\documentclass[%
 aip,
 amsmath,amssymb,
 reprint,%
]{revtex4-1}

\usepackage{graphicx}
\usepackage{dcolumn}
\usepackage{bm}

\usepackage[utf8]{inputenc}
\usepackage[T1]{fontenc}
\usepackage{mathptmx}
\usepackage{etoolbox}
\usepackage{color}

\makeatletter
\def\@email#1#2{%
 \endgroup
 \patchcmd{\titleblock@produce}
  {\frontmatter@RRAPformat}
  {\frontmatter@RRAPformat{\produce@RRAP{*#1\href{mailto:#2}{#2}}}\frontmatter@RRAPformat}
  {}{}
}%
\makeatother
\begin{document}

\preprint{AIP/123-QED}

\title{How important is the dielectric constant in water modeling?
Evaluation of the performance of the TIP4P/$\varepsilon$ force field and its compatibility with the Joung-Cheatham NaCl model}
\author{{\L}ukasz Baran}
\email{lukasz.baran@mail.umcs.pl}
\affiliation{ Department of Theoretical Chemistry, Institute of Chemical Sciences, Faculty of Chemistry, Maria Curie-Sklodowska University in Lublin, Lublin, Poland.  }
\affiliation{Departamento de Qu\'{\i}mica F\'{\i}sica, Facultad de Ciencias
Qu\'{\i}micas, Universidad Complutense, Madrid, 28040, Spain.}

\author{Cosmin A. Dicu-Gohoreanu}%
\affiliation{Instituto de Química Física Blas Cabrera, CSIC
, C/Serrano 119, 28006 Madrid, Spain}

\author{Luis G. MacDowell}
\affiliation{Departamento de Qu\'{\i}mica F\'{\i}sica, Facultad de Ciencias
Qu\'{\i}micas, Universidad Complutense, Madrid, 28040, Spain.}

\date{\today}

\begin{abstract}
Efficient large-scale computer simulations of aqueous solutions require the use of accurate but simple empirical force fields for water. However, the complexity of these systems evidences the difficulties in describing solution properties without due account of polarization. {\color{black} Different} strategies to remedy this problem are parametrizing water force fields to the dielectric constant or charge scaling of solvated ions. In this work, we compare results from TIP4P/$\varepsilon$ and OPC models, which are parametrized to predict the dielectric constant, with results from TIP4P/2005, which is closer in spirit to the charge scaling strategy. The performance of the models is rated according to the Vega-Abascal benchmark. Our results show that TIP4P/$\varepsilon$ and TIP4P/2005 perform equally well, with the OPC model lying significantly behind. TIP4P/$\varepsilon$ can predict bulk phase properties (transport properties, thermal expansion coefficients, densities) of both liquid water and ice polymorphs, but also surface tensions, with an accuracy very similar to TIP4P/2005, while performing very well for dielectric constants over a wide range of pressures and temperatures. On the other hand, TIP4P/2005 provides a better description of phase boundaries, including liquid-vapor and freezing transitions. However, the accurate prediction of dielectric constants allows TIP4P/$\varepsilon$ to describe {\color{black} densities of NaCl solutions for models  parametrized to their crystal and melt properties only.} This is achieved without the need to rescale charges, modify the Lorentz-Berthelot rule or tune the ion's Lennard-Jones parameters. Our findings hinge on the significance of dielectric constants as a target property and {\color{black} show that a robust parametrization can be achieved without invoking} the concept of charge scaling.
\end{abstract}

\maketitle

\section{Introduction}

Water is essential for the existence of life  on Earth. A simple molecule, yet exhibiting numerous anomalies, such as a negative expansion coefficient below 4~$^\circ$C, a sharp increase of response functions upon cooling, a decrease of viscosity with the increase in pressure, and many others \cite{gallo16}.

This unique behavior makes water a perfect testbed for the study of a wide range of fundamental physical principles in statistical mechanics. 
Therefore, great efforts have been devoted to the development of accurate and meaningful force fields ever since the advent of computer simulations \cite{ouyang15}.

On physical grounds, it is clear that the design of an all-purpose model which can adapt depending on the environment requires dealing with the polarizability issue, if not exactly, at least approximately. Unfortunately, polarizable models such as the BK3 \cite{kiss13}, HBP \cite{jiang16}, and iAMOEBA \cite{wang13} give only a modest improvement, but their computational cost makes them inefficient for large system sizes. On the other hand, the very active line of research based on machine learned potentials can do no better than the underlying approximations \cite{zhang18}, and some recent studies for state of the art density functionals provide very disappointing results \cite{montero23, montero24}. Further improvement along this line requires  a training set obtained from expensive  ab initio calculations dealing accurately with electronic correlations, but alas, neural networks appear to exhibit here significant limitations in modelling the ab initio potential energy surface for a wide range of different chemical environments \cite{zhai23}.

Due to these difficulties, model development is still often directed to the design of simple and tractable  force fields, based on intuition and basic principles of  chemistry. As a result, one often neglects polarizability altogether and   describes the average molecular polarity with a simple distribution of fixed point charges embedded into a spherical core. This is the case of  the celebrated SPC model \cite{berendsen87}. Here, two positive partial charges sitting on hydrogen positions are neutralized by a negative charge located at the center of a simple Lennard-Jones site, which accounts for the repulsion and dispersion of the  oxygen atom. Alternatively, the negative charge can be shifted along the bisector axis of the HOH angle, as proposed by Bernal-Fowler  already in 1933 \cite{bernal33}, leading to the so-called TIP4P family of water models.\cite{jorgensen83} 

Despite their simplicity, SPC and TIP4P models do a rather good job at describing the qualitative behavior of liquid water at ambient conditions, but the TIP4P family has been shown to be more robust and remains qualitatively correct over a much wider range of thermodynamic conditions \cite{sanz04,abascal07c}. Particularly, the TIP4P/2005 model provides a good qualitative description of the phase diagram, including the  coexistence between several solid phases, as well as a quantitative description of the temperature of maximum density, the orthobaric liquid densities, molar volumes and surface tensions \cite{abascal05b,abascal07c}. 

Of course, neglecting polarizability comes at a price. For the particular case of the TIP4P/2005 model, the accuracy of liquid phase densities is at the cost of a rather poor description of the dielectric constant, which is about 30\% too low compared to experiments \cite{vega11}. Alternatively, it often occurs that models providing a good description of the dielectric constant, such as SPC/E or TIP3P \cite{berendsen87,jorgensen83}, yield a qualitatively incorrect description of the phase diagram \cite{vega11}. 

A simple way out of this problem was suggested by Leontyev and Stuchebrukhov some time ago \cite{leontyev09, leontyev11}. These authors argued that the electric field generated by point charges on a molecule is screened by the electronic polarizability of the surrounding molecules. Therefore, the {\em "effective"} charges, $q_{\rm eff}$  that are required to reproduce an accurate force field  are smaller than the {\em true} point charges of the molecule, $q_{\rm true}$ by a factor equal to the high frequency refractive index, $n_{\infty}$, i.e. $q_{\rm eff}=\frac{q_{\rm true}}{n_{\infty}}$. Consequently, models predicting a good force field will yield too low a molecular dipole moment when measured in terms of the effective charge. However,  assuming the true charge is a factor $n_{\infty}$ larger than the effective charge, and noting water's refractive index is  ca. $n_{\infty}=1.33$, brings the dipole moment of usual force fields   ($\approx 2.2-2.4$~D) in close agreement with results of ab initio calculations  ($\approx 3.0$~D).

This appealing idea has received great attention  and has become known as the  {\em scaled charge} paradigm \cite{kann14,kirby19,jorge19,zeron19,han23,weldon23,blazquez23a}. As another side of the same coin, some authors have  stressed the need to make a distinction between the potential energy surface, as governed by $q_{\rm eff}$, and the so called {\em dipole moment surface}, which is dictated by  $q_{true}$\cite{vega15}.  This concept can be useful to bring predicted dielectric constants into agreement with experiment \cite{aragones11b,han23}, but must be exercised with great caution. Indeed,  these two surfaces could be different but are definitively not  independent. Linear response theory implies that a molecule cannot tell the difference between an external electric field (as dictated by a test charge) or  that emanating from a neighboring molecule within the system (as dictated by the effective charge). However, in the scaled charge paradigm the dipolar response would be dictated by the true dipole in the former case and by the scaled dipole in the latter.  As a result, arbitrarily scaling  the molecular dipole  to bring the dielectric constant into agreement with experiments is likely to violate the 
dissipation-fluctuation theorem.

Be as it may, the charge scaling concept serves to increase the number of liquid state properties that can be predicted by non-polarizable force fields, and optimized models such as TIP4P/2005 can be considered as very robust tools for the prediction of pure water properties. However, the lack of polarization will result in poor performance in cases where environments with widely differing structures are encountered  in the same system \cite{abascal07c}.  

One such important situation is when two or more phases need to be considered simultaneously, as in phase coexistence. Indeed, it has been widely documented that it is not possible to predict accurate liquid water properties while providing a good estimate of the melting point \cite{abascal07c}. Similarly, point charge models provide usually rather poor vapor pressures, even when including ad hoc corrections for the  free energy of polarization \cite{berendsen87,vega06b}. 

A second very significant case involves electrolyte solutions, where ions can create very strong local electric fields and polarization becomes then a major issue. Indeed, for a single isolated ion of charge $q$, the free energy of solvation is proportional to $q^2/\varepsilon$ \cite{israelachvili11}, so one can anticipate that the lack of an accurate dielectric constant will result in a poor description of the solution. Not surprisingly, attempts to reproduce {\color{black} properties of} electrolyte solutions using the expected nominal charge  have met only limited success  \cite{joung08, yagasaki20, smith94, fuentes16}.  

The scaled charge concept comes here at the rescue, under the assumption that also the nominal charge of the ions needs to be scaled \cite{kann14}. Indeed, several authors have recently exploited this idea \cite{kann14,kirby19,jorge19,zeron19,han23,blazquez23b}, and showed that some of the TIP4P models with poor predictions of the dielectric constant can provide rather accurate predictions of a number of solution properties by scaling the nominal charges by a factor of ca. $1/n_{\infty}=0.75$ \cite{zeron19, blazquez22, trejos23,blazquez23a}. Unfortunately, predictions here are far less robust than for pure water, and different scaling factors need to be considered depending on whether one seeks for accurate predictions of bulk thermodynamic properties, surface tensions, or transport properties \cite{blazquez23a, breton20}. Moreover, the improvement of the solution properties comes at the cost of significantly deteriorating the description of the melt and the crystal, which also results in poor predictions of the electrolyte's solubility.  

For this reason, some authors have sought for an improved description of solution properties by using water's dielectric constant as a major target property in the model parametrization \cite{izadi14,fuentes14} 

This is the case of the OPC model\cite{izadi14}, which has been recently recommended for use with the Amber package \cite{tian20}. Unfortunately, the improved description of the dielectric constant in this model is at the cost of using the location of the hydrogen sites as fitting parameters. This can provide extra flexibility for the parametrization, but will obviously ruin the molecule's moment of inertia, which is one of the very few molecular properties that we can safely constrain from the outset. Of course, the dynamics will suffer from such a choice, and recent work has shown that the increased parameter space chosen is not sufficient to improve other models such as TIP4P/2005 \cite{sedano24}. On the other hand,  TIP4P/$\varepsilon$, a model developed by Fuentes-Azcatl \textit{et al.} at about the same time \cite{fuentes14},  reproduces experimental dielectric constants over a wide range of thermodynamic conditions, with only small variations of the TIP4P/2005 charges and geometry of the parent model. As a result, it can also predict accurately the location of the temperature of maximum density. The simultaneous prediction of these two properties is rather uncommon in point charge force fields and appears to suggest a good performance of solution properties. 

Based on the scaled charge paradigm, however, it would appear that using dielectric constants as a target property is unsafe and will result in a poor description of the\textit{"true"} force field \cite{sedano24}. According to this argument, one would then expect that the improved performance will be at the cost of spoiling other properties, as it appears to be the case for the OPC model \cite{sedano24}.

In this work, we assess the performance of the TIP4P/$\varepsilon$ water model for a wide range of different properties and compare its performance with OPC and TIP4P/2005 using the well known  benchmark proposed by Abascal and Vega (VA) \cite{vega11}. The model is then tested for transferability by studying solution properties as described by the Joung-Cheatham electrolyte force field \cite{joung08}, which was not specifically parametrized for TIP4P/$\varepsilon$. Our results show that TIP4P/$\varepsilon$ performs essentially as well as TIP4P/2005 for pure water and that it remains a transferable and robust model for the study of electrolyte solutions without additional parametrization. This {\color{black} shows that a robust parametrization can be achieved without invoking} the appealing concept of charge scaling.

\section{Overview of charge scaling}

Under the scaled charge paradigm, we acknowledge that the fixed net charges of ions and point polarized solvent molecules do not properly account for electronic polarization effects.

In order to account for this problem effectively, while retaining the convenient point charge interactions, we assume that the electronic degrees of freedom may be described effectively as a dielectric continuum embedding both the partial charges of molecules and net charges of ions  in the system.

Such a continuum description makes sense provided the distance between the interacting charges is large compared with the typical distances between the polarizable solvent molecules. Charge scaling is therefore the exact long wavelength result for the interaction between ions in a solvent. Assuming that interactions between nearby ions or solvent molecules are mediated by a dielectric continuum is likely to be significantly less accurate, however. 

Let us ignore this limitation for the time being, and assume the electrostatic approximation  is adequate for the atomic nuclei in their fixed positions at a given instant. Then, from Maxwell's equations, the electric field is related to the charge distribution in the system as:
\begin{equation}\label{eq:max}
\nabla \cdot {\bf E} = 4\pi \rho
\end{equation}
where $\rho$ is the total charge density. This is made of the fixed point charges of the model, $\rho_f$, and implicit charges that result from the polarization, ${\bf P}$, of the electronic continuum background, $\rho_p = -\nabla \cdot {\bf P}$.

Assuming an isotropic, local and linear response of the dielectric background, the polarization is ${\bf P}=\chi {\bf E}$, where $\chi$ is the background's susceptibility. In as much as the nuclei are fixed, and the polarization is only due to the electronic degrees of freedom, we can assume  $\chi$ is dictated by the static electronic polarization of the system. In practice, electronic degrees of freedom are unresponsive below  {\color{black} an optical frequency, $\omega_{VIS}$}    \cite{parsegian05,israelachvili11}. Therefore, the response of the medium is dictated by the system's index of refraction in the visible, $n_{\infty}$, {\color{black} which is related to the high frequency dielectric constant as $\epsilon(\omega_{VIS})=n_{\infty}^2$} .

Under these assumptions, the energy between interacting charges  is given readily by the laws of electrostatics in a polarizable medium, with $n_{\infty}^2$ playing the role of the medium's dielectric constant:
\begin{equation}\label{eq:cs}
	u_{ij} = \frac{q_i q_j}{n_{\infty}^2 r_{ij}}
\end{equation}
{\color{black} Notice that we have taken care to write here the screening felt by charges in terms of the solvent's refraction index in order to make explicit that it is only the electronic degrees of freedom that participate, i.e., the screening here does not account for the low frequency permanent dipole fluctuations of the solvent, and it is more akin to $\epsilon(\omega_{VIS})$ than to the static dielectric constant $\epsilon(\omega=0)$.}

Eq.(\ref{eq:cs}) provides the rationale for charge scaling. Since the pair energy is dictated by the scaled charges, $q_i/n_{\infty}$, and the dielectric constant measures fluctuations of the actual charges, $q_i$, it would appear to be unwise to use dielectric constants as a target property in force field modeling.

However, at a finer level of description, we can still approximate the medium as a dielectric continuum, but take into account that the polarization at a point is non-local, and depends also on the electric field at nearby points:
\begin{equation}\label{eq:nlpol}
 {\bf P({\bf r})} = \int d {\bf r}' \chi({\bf r}-\bf{r}') {\bf E}({\bf r}')
\end{equation}
where $\chi({\bf r}-\bf{r}')$ is now the non-local susceptibility of the homogeneous medium.

Assuming an isotropic response in Eq.(\ref{eq:nlpol}), Eq.(\ref{eq:max}) can be solved in Fourier space, and eventually leads to a Fourier mode decomposition of the interaction potential as given by:
\begin{equation}\label{eq:maxk}
   \phi(k) = \frac{4\pi \rho_f(k)}{n_{\infty}^2(k) k^2}
\end{equation}
{\color{black} 
where the non-local electronic response $\epsilon(k)=n_{\infty}^2(k)$ is now given in terms of the wave-vector dependent susceptibility as $n_{\infty}^2(k)=1+4\pi\chi(k)$.

This shows implicitly that the extent of screening felt by the molecules is position dependent. 
In order to illustrate this explicitly, a model for the complex refraction index, $n_{\infty}(k)$ is needed. For simple models of water and dipolar hard spheres,\cite{stell81,trokhymchuk93,bopp96,bopp98,seyedi19,becker25} simulations and integral equation theory show that the wave-vector dependent dielectric constant has a rather complex behavior, and exhibits at least two poles at finite wave-vectors. However, those calculations refer to the static dielectric constant, which is dominated by the fluctuations due to permanent dipole moments of the solvent.\cite{stell81} Here, we are only interested in the contribution of electronic fluctuations for frozen realizations of the nuclear positions, which is a problem far less well understood.  In the absence of a suitable theoretical framework, we suggest here a minimal model, meant to illustrate how the distance dependent screening could emerge from the wave-vector dependent refractive index. For this purpose, we consider that two very nearby charges in a dilute gas effectively feel no screening, while two very distant charges will be screened as in a continuum.  Assuming an even power dependence borrowed from statistical mechanics and field theory,\cite{hansen86,trokhymchuk93,bopp96,seyedi19} a simple interpolating formula between these two limits is:
\begin{equation}\label{eq:model}
 n_{\infty}^2(k) = 1 + \frac{n_{\infty}^2-1}{1 + (k \xi)^2}
\end{equation}
where $\xi$ is a correlation length in the order of the solvent molecular spacing. This equation resembles the Random Phase Approximation for the one component plasma, but with a cutoff to avoid the low wave-vector divergence that is typical in metals.\cite{hansen86}

}

Fourier transforming Eq.(\ref{eq:maxk}), using Eq.({\ref{eq:model}) for the polarization response yields an effective pair potential for the electronic background mediated interactions (cf. \cite{mondal24}):
\begin{equation}\label{eq:screening}
  u_{ij}(r) = \frac{q_i q_j}{n_{\infty}^2 r} \left( 1 + ( n_{\infty}^2-1 ) e^{- \frac{r}{\lambda} } \right) 
\end{equation}
This result smoothly interpolates between Coulomb's law in a vacuum and a screened Coulomb's law for charges in a polarizable background. The crossover takes place at a microscopic length scale of $\lambda=\xi/n_{\infty}$. 
This result might be a mere caricature of the complex distance dependent screening in water \cite{bopp96,bopp98,berthoumieux18}, but helps to show how the simplest account of non-locality can lead to deviations from the continuum approximation.
 Interestingly, Eq.(\ref{eq:screening}) can be also interpreted as describing a plain Coulomb law with an $r$ dependent permittivity of $n^2(r)=n^2_{\infty}/(1 + (n^2_{\infty}-1) e^{-r/\lambda})$, which smoothly switches from unscreened interactions to the expected long wave-length limit for large $r$. The crucial issue here is what is the extent of charge screening at the typical distance of the closest approach between the molecules, say, $\sigma$. i.e., whether $e^{-\sigma/\lambda}$ is already sufficiently small for full screening to have set on. Since we expect $\xi$ on the scale of $\sigma$, this is not likely to be the general situation.  This is in line with recent ab initio calculations between ions dissolved in argon, which show that the screening at contact distances is not quite that achieved at large distances, but at least, definitively  much larger than that expected for ions in vacuum \cite{kostal23}. This means that charge screening is not fully warranted at contact distances, but could be a  better approximation than assuming charges interacting in vacuum.

\section{Methods}

\subsection{Dielectric constant}
To determine the dielectric constant of both pure water and NaCl electrolyte solutions we used the following formula:

\begin{equation}\label{eq:epsilon}
    \varepsilon=1+\frac{4\pi}{3 k_B T V}\left<M_x^2+M_y^2+M_z^2\right>
\end{equation}

\noindent where $k_B$ is the Boltzmann constant, $T$ and $V$ the temperature and volume, respectively, and $M_j$ is the $j-$th component of the total dipole moment. The simulations performed in the $NpT$ ensemble lasted at least $30$~ns in order to gather sufficient statistics. 

\subsection{Melting point}
The melting point of ice Ih was estimated using the direct coexistence method \cite{garcia06} where two phases at coexistence are brought together across an interface. Based on the time evolution of a number of liquid molecules, identified by the CHILL+ algorithm \cite{nguyen15}, we have estimated the melting/freezing rates. Afterwards, these rates were plotted as a function of temperature and the melting point was located by interpolation of results to zero rate. The simulations performed in the $Np_\perp AT$ ensemble lasted up to $60$~ns to unambiguously determine the melting/freezing modes. 

\subsection{Infrared spectrum}

The infrared spectra of rigid point charge models are estimated following a procedure suggested by Skinner and collaborators \cite{corcelli04,auer07,auer08,gruenbaum13}. In this method, the vibrational frequencies of water clusters calculated from ab initio calculations are correlated to the electric fields generated by a rigid point charge model. In this way a spectroscopic map is created which allows one to predict the infrared spectra from the electric field generated by a force field \cite{takayama23}. Here, we used the time averaging approximation, whereby the instantaneous vibrational frequencies as obtained from the spectroscopic map are averaged over a period of the order of the librational timescale. The spectrum is then predicted as the distribution of such frequencies over the full simulation, according to  \cite{auer07}:
\begin{equation}
    I(\omega) = \langle \delta(\omega - \omega_T) \rangle
\end{equation}
where $\omega_T$ is the time averaged frequencies. Here we focus on the OD tension of a solution of deuterated water (HDO) in H2O, under the assumption that any one single molecule of the system is representative of the electric field felt by the solvated HDO molecule \cite{takayama24}. This choice is convenient, because the OD chromophore does not couple to the neighboring OH chromophores, due to the large separation of the frequencies. As a result, we need not embark on the complications of inter or intra molecular coupling \cite{auer07,auer08,gruenbaum13}. In order to estimate instantaneous frequencies to each molecular environment as sampled from the simulations, we use the OD stretch map parametrized for the TIP4P model \cite{gruenbaum13}. A number of different studies point to the robustness of this mapping to change of model potentials and local environmental conditions \cite{gruenbaum13,takayama23,yamaguchi23,takayama24} Instantaneous frequencies are time averaged over 100~ps, as recommended in Ref.\cite{kananenka18} Averages are collected from NVE simulations with 256 molecules, using a time step of 0.5~fs in order to guarantee an accurate calculation of the high frequency dynamics \cite{dicu24}.

\subsection{Transport properties}

Self-diffusion coefficients were calculated using the Einstein relation, involving the calculation of the mean-squared displacement (MSD) of individual water molecules

\begin{equation}
    \left<\Delta r^2(t)\right>=\left<(\mathbf{r}(t)-\mathbf{r}(0))^2\right>
\end{equation}

\noindent where $\mathbf{r}(t)$ is the position of a water molecule at time $t$, and the triangular brackets denote a thermal average over all time origins. 
The diffusion coefficient is then related to the slope of the MSD as $\left<\Delta r^2(t)\right>=6D_{PBC}t$. To account for the finite-size effects we have included the Yeh-Hummer (YH) correction \cite{Yeh04}, defined as 

\begin{equation}
    D_{PBC}=D_0-\frac{ 2.837297 k_BT}{6\pi\eta L}
\end{equation}

\noindent where $\eta$ is the viscosity, and $L$ the length of the cubic simulation box.

The shear viscosity was calculated from the Green-Kubo formula:

\begin{equation}
    G_{\alpha\beta}=\frac{V}{k_BT}\left<\sigma_{\alpha\beta}(t)\sigma_{\alpha\beta}(0)\right >
\end{equation}

\noindent where $\sigma_{\alpha\beta}$, with  $\alpha,\beta=x,y,z$ represent the components of the stress tensor. To improve the statistics, we exploit both the off-diagonal and the diagonal components of the stress tensor. This is allowed provided the off-diagonal elements are weighted by the adequate factors \cite{daivis94}. Taking this into account, the shear viscosity is calculated as: 

\begin{equation}
    \eta=\int_0^\infty G_{\eta}(t)dt
\end{equation}

\noindent where $G_{\alpha\beta}=\frac{1}{6}[G_{xx}+G_{yy}+G_{zz}
+\frac{3}{4}(G_{xy}+G_{xz}+G_{yz})]$

The simulation scheme for both transport properties involved two steps. In the first one, auxiliary $NpT$ simulations were performed for $10$~ns, allowing us to estimate the average box size and density of the fluid. Afterwards, the system was rescaled to the average dimensions estimated before and production runs lasting up to $30$~ns in the $NVT$ ensemble were launched. This avoids possible spurious effects of the barostat in the dynamics of the system.

\subsection{Properties of ice polymorphs}

The density of ice II, V and VI at selected conditions was calculated using a Monte Carlo code used previously in the determination of the phase diagram of ice and the parametrization of TIP4P/2005 and TIP4P/Ice models\cite{sanz04,abascal05,abascal05b}. Details of the code may be found in Ref.\cite{baran24b}. Details on the preparation of initial configurations of the ice polymorphs are described in Ref.\cite{sanz04,macdowell04b} The dielectric constant of ice Ih was calculated using a ring rotation algorithm as described in Ref.\cite{rick03,rick05,macdowell10}

\subsection{Molecular dynamics setup}

\begin{table} [h!]
    \centering
    \begin{tabular}{ccc}
    \hline
    \hline
        System & Number of molecules & $L_x\times L_y\times L_z$ ~(\AA$^3$)\\
        Bulk ice & 1280  & $36\times32\times36$ \\
        Bulk water & 2220 & $40\times40\times40$\\
        Melting point & 10240 & $90\times63\times59$  \\
        Electrolyte solutions & 2220 + $c\cdot 40$ salt pairs & $40\times40\times40$\\
    \hline
    \hline
    \end{tabular}
    \caption{Different system sizes considered for the evaluation 
    of properties in this paper. $c$ depicts the molality concentration (mol kg$^{-1}$) of the electrolyte solution.}
    \label{tab:sizes}
\end{table}

The molecular dynamics simulations are performed using the LAMMPS package \cite{lammps22}. Water was modeled using the TIP4P/$\varepsilon$ force field \cite{fuentes14} and for selected properties we have also performed calculations for the TIP4P/2005 model. The number of molecules and box sizes varied in different simulation sets and are summarized in Table~\ref{tab:sizes}. 
Trajectories are evolved using the velocity-Verlet integrator using a time step of $1$~fs. For pure water, the equations of motion were solved with the SHAKE algorithm, with the dummy atom defined implicitly according to the \textit{pair\_style lj/cut/tip4p/cut} command. Unfortunately, this command does not allow for the correct calculation of electrostatic energy in the presence of ions. Therefore, in the case of solutions, a slightly less efficient quaternion-based rigid body dynamics was used   \cite{kamberaj05}. Both the temperature and pressure were maintained using a three chains Nos{\'e}-Hoover algorithm with a damping factor 
$\tau=2$~ps. Tail corrections were included. Dispersion interactions were truncated at the separation distance of $10$~\AA. Long-range electrostatics were computed using the particle–particle particle-mesh method \cite{darden93}. The charge structure factors were evaluated with the fourth-order interpolation scheme and a grid spacing of 1~\AA.

\subsection{Force field parameters}

In this work, we have performed simulations using the TIP4P/$\varepsilon$ water model \cite{fuentes14} which has been compared to the two other non-polarizable force fields -  TIP4P/2005 \cite{abascal05} and OPC \cite{izadi14}. The entire set of parameters, required for launching the simulations is summarized in Table~\ref{tab:models}. 

\begin{table}[h!]
    \centering
    \resizebox{\linewidth}{!}{
    \begin{tabular}{ccccccccc}
        Model & Charge (e) & $\sigma$~(\AA) & $\varepsilon$~(kJ/mol)  & $d_{OH}$~(\AA) & $d_{OM}$~(\AA) & $\theta$~$(^\circ)$ & $\mu$~(D) & $Q_T$  \\
        \hline
        \hline
        \multicolumn{9}{l}{TIP4P/$\varepsilon$} \\
        \hline
        O & 0. & 3.165 & 0.7732 & 0.9572 & 0.105 & 104.52  & 2.4345 & 2.174  \\
        H & 0.527  \\
        M &-1.054  \\
        \hline
        \multicolumn{9}{l}{TIP4P/2005} \\
        \hline
        O & 0. & 3.1589 & 0.7749 & 0.9572 & 0.1546 & 104.52  & 2.305 & 2.297  \\
        H & 0.5564  \\
        M &-1.1128  \\
        \hline
        \multicolumn{9}{l}{OPC} \\
        \hline
        O & 0. & 3.16655 & 0.89038 & 0.8724 & 0.1594 & 103.6  & 2.48 & 2.3  \\
        H & 0.6791  \\
        M &-1.3582  \\
        \hline
        \hline
    \end{tabular}
    }
    \caption{Comparison of the parameters for the TIP4P/$\varepsilon$ \cite{fuentes14}, TIP4P/2005 \cite{abascal05}, and OPC \cite{izadi14} water models.}
    \label{tab:models}
\end{table}

\section{Results and Discussion}

\subsection{Dielectric constant}

\begin{figure}[h!]
    \centering
    \includegraphics[width=0.8\linewidth]{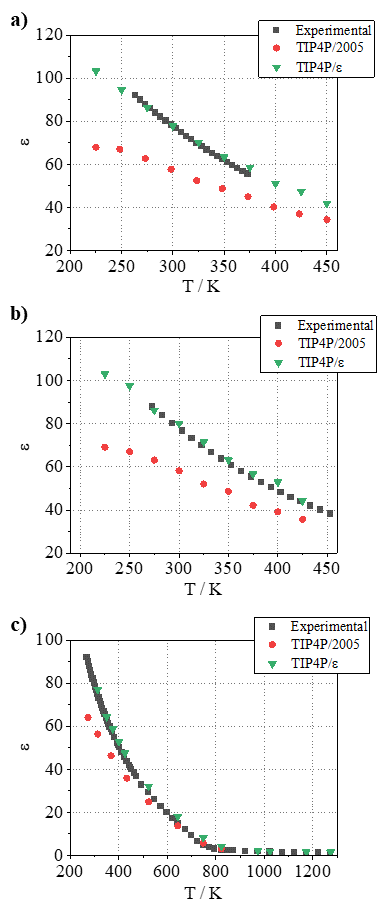}
    \caption{Static dielectric constant as a function of temperature 
    at $p=1$~bar (a), $p=p_{sat}$ (b), and $p=500$~bar (c). Experimental data are taken from Ref.~\cite{fernandez95, uematsu80}.}
    \label{fig:dielectric}
\end{figure}

Let us begin with the discussion related to the static dielectric constant of the TIP4P/$\varepsilon$ water model. In the initial paper, Fuentes-Azcatl \textit{et al.} \cite{fuentes14} argued that Lennard-Jones and Coulombic interactions behave  independently. This allows us to choose the charge distribution to reproduce the dielectric constant and then determine the LJ parameters to reproduce the temperature of maximum density (TMD) and other properties. In fact, they have shown that this approach was successful and the developed water model could reproduce experimental values of both properties. 

To check this, we have complemented the results of Fuentes-Azcatl and Alejandre over  a wider range of thermodynamic conditions. This includes two isobars at 1 and 500~bar, as well as results along the saturation curve of water.  The results for both the TIP4P/2005 and TIP4P/$\varepsilon$ model are shown in Figure~\ref{fig:dielectric}. It is evident that the TIP4P/$\varepsilon$ model reproduces experimental data very accurately, contrary to the TIP4P/2005 model, which predicts dielectric constants that deviate from the experimental values by more than 25 \%.

At first thought, one could suspect that such a large increase of the dielectric constant can be achieved by artificially increasing the molecular dipole with too large point charges. This would then seriously compromise the accuracy of the electric field generated by the molecules, thus spoiling the actual force field. Accordingly, the good agreement of dielectric constants would  be at the cost of seriously deteriorating prediction of the remaining properties. Actually, inspection of the model parameters (cf. Table~\ref{tab:models}) shows that the  TIP4P/$\varepsilon$ has point charges that are just  5\% smaller than those of the TIP4P/2005. The crucial difference between the models appears to be the placement of the dummy particle, which is located much closer to the oxygen site in TIP4P/$\varepsilon$. The overall effect of this change is to increase the single molecule dipole moment by merely 5\%. Accordingly, the large increase in the dielectric constant cannot be attributed to a crude scaling of the dipole.

This can be seen by recasting the expression for the dielectric constant, Eq.\ref{eq:epsilon} as:
\begin{equation}
    \varepsilon=1+\frac{4\pi\rho}{3 k_B T} \mu^2 g_K
\end{equation}
where $\mu$ is the molecular dipole moment, and the Kirkwoods factor, $g_K$, is a measure of the static correlation of the molecular orientation as given
by a unit vector along the direction of the molecular dipole, ${\bf u_i}$:
\begin{equation}
    g_K = \frac{1}{N} \left \langle \sum_{i,j} {\bf u_i}\cdot{\bf u_j} \right \rangle
\end{equation}
Given that the dipoles in TIP4P/2005 and TIP4P/$\varepsilon$ are both directed along the HOH bisector, and that their dipole moments are very similar, it follows that much of the change in the dielectric constant must be related to significant differences in the orientational correlations. 

For liquid water at ambient temperature, we find that TIP4P/2005 predicts $g_K=3.2$, while TIP4P/$\varepsilon$ predicts $g_K=3.85$, implying stronger orientation correlations in the latter model. 

A very stringent test for non-polarizable models is to check the drop of the dielectric constant upon freezing water. In experiments, the dielectric constant at the melting point of ice increases from 78 for ice cold water, to 107 for ice Ih. This large increase has never been predicted for any of the point charge models. In fact, these models, on the contrary, predict a significant drop of the dielectric constant on freezing \cite{aragones09,macdowell10}. 
TIP4P/$\varepsilon$ is no exception. Our calculations show that for ice at 273~K, the dielectric constant decreases down to 43, due to a very large reduction of the Kirkwood factor, which drops to $g_K=2.0$. On the other hand, TIP4P/2005 exhibits a dielectric constant of $\epsilon=46$, and now achieves a Kirkwood factor $g_K=2.46$ larger than that of TIP4P/$\varepsilon$. Whence, although the two models exhibit similar dielectric constants at this temperature (where ice Ih is actually a metastable phase for both models), we see that the behavior of TIP4P/2005 is qualitatively somewhat better than that of TIP4P/$\varepsilon$, as it predicts a much smaller drop of the dielectric constant upon freezing. In fact, the behavior of TIP4P/$\varepsilon$ somewhat resembles that of the SPC/E model, where the Kirkwood factor decreases from 3.68 for water at ambient temperature to barely 1.74 for ice \cite{aragones09}. This similarity is reasonable, since TIP4P/$\varepsilon$ has the dummy site displaced from the oxygen by a much smaller amount than  TIP4P/2005, and is, therefore, more closely related to SPC/E. If this analogy holds, TIP4P/$\varepsilon$ is not a reliable model for the study of ice polymorphs. This statement will be assessed later on.

\subsection{Infra Red spectrum}

\begin{figure}[h!]
    \centering
    \includegraphics[width=\linewidth]{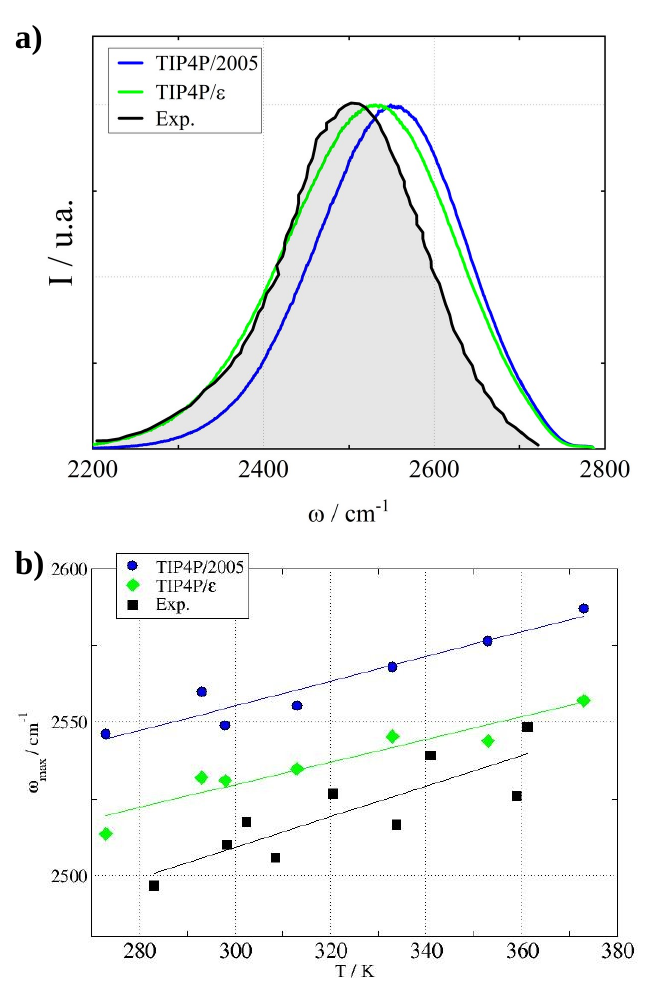}
    \caption{a) Infrared spectra of the OD stretch of an HDO molecule dissolved in H2O as predicted by TIP4P/2005 (blue) and TIP4P/$\varepsilon$ models. Results are compared with the experimental data as reported in Ref.\cite{corcelli04}. b) Dependence of the OD stretch frequency with temperature, compared with experimental results as compiled in Ref.\cite{corcelli05}. The straight lines are linear fits mainly to serve as a guide to the eye. }
    \label{fig:infrared}
\end{figure}

Since TIP4P/2005 and TIP4P/$\epsilon$ seem to produce rather different charge distributions, it is instructive to move on to study high frequency dipolar fluctuations of condensed water, which are very sensitive to the local electric fields generated on the water atoms \cite{corcelli04,corcelli05,auer07}. The OH stretch band of the Infrared spectrum is a particularly convenient probe. However, in order to avoid complications due to inter and intra molecular couplings, here we consider the OD stretch of an isolated HDO molecule dissolved in H$_2$O, which behaves as an uncoupled chromophore \cite{corcelli04,corcelli05,kananenka18,dicu24}.  

Of course, rigid models do not exhibit an explicit OD stretch at all, but it is possible to produce a synthetic spectrum from the study of the local electric fields generated on the Deuterated atoms \cite{corcelli04,auer07,auer08,gruenbaum13}. Although this is an empirical approach, which maps ab initio results for the OD stretch to the electric field of a reference TIP4P model, it has been argued that the mapping is transferable and can be exploited to assess the performance of different force fields \cite{takayama23,yamaguchi23,takayama24}.

The results of the OD spectrum are shown in Figure~\ref{fig:infrared}. From panel a) we see that TIP4P/$\epsilon$ does a much better job than TIP4P/2005 at describing the low frequency side of the OD band, but has a significantly larger band width, and does therefore not reproduce the high frequency side of the band accurately. On the other hand, the spectrum predicted by TIP4P/2005 has a band center that is some decades of cm$^{-1}$ blue shifted with respect to experiment but shows a band width closer to the experimental spectrum.

Both models do a good job at reproducing the blue shift of the OD stretch as temperature increases, with TIP4P/$\epsilon$ yielding a band center in closer agreement with the experiment. However, the slope of the band center as a function of temperature is given more accurately in the TIP4P/2005 model.

This assessment leaves the discussion of the accuracy of the models somewhat undecided, and we resort to a discussion of other thermodynamic properties in the next few sections.

\subsection{Densities}
\begin{figure}[h!]
    \centering
    \includegraphics[width=0.8\linewidth]{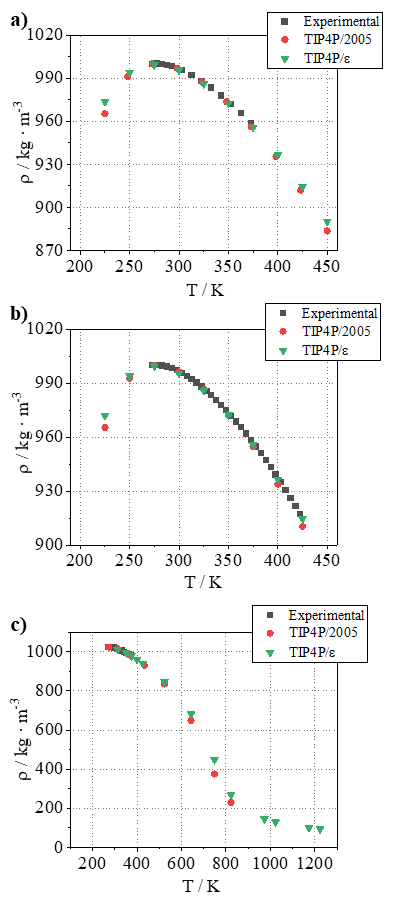}
    \caption{Density of water as a function of temperature 
    at $p=1$~bar a), $p=p_{sat}$ b), and $p=500$~bar c). Experimental data is taken from Ref.~\cite{kell75}.}
    \label{fig:densities}
\end{figure}

The next property that has been calculated was the water's density as a function of temperature for three different pressures. The results are shown in Figure~\ref{fig:densities}. It is clear that both TIP4P/2005 and TIP4P$\varepsilon$ water models are quantitatively reproducing experimental results in a wide range of thermodynamic conditions. We therefore corroborate the observation of Azcatl and Alejandre that both the dielectric constant and TMD can be accurately reproduced by the same model. The question then remains to what extent are other predictions affected by this choice of parameters.

\subsection{Melting point}

\begin{figure}[h!]
    \centering
    \includegraphics[width=\linewidth]{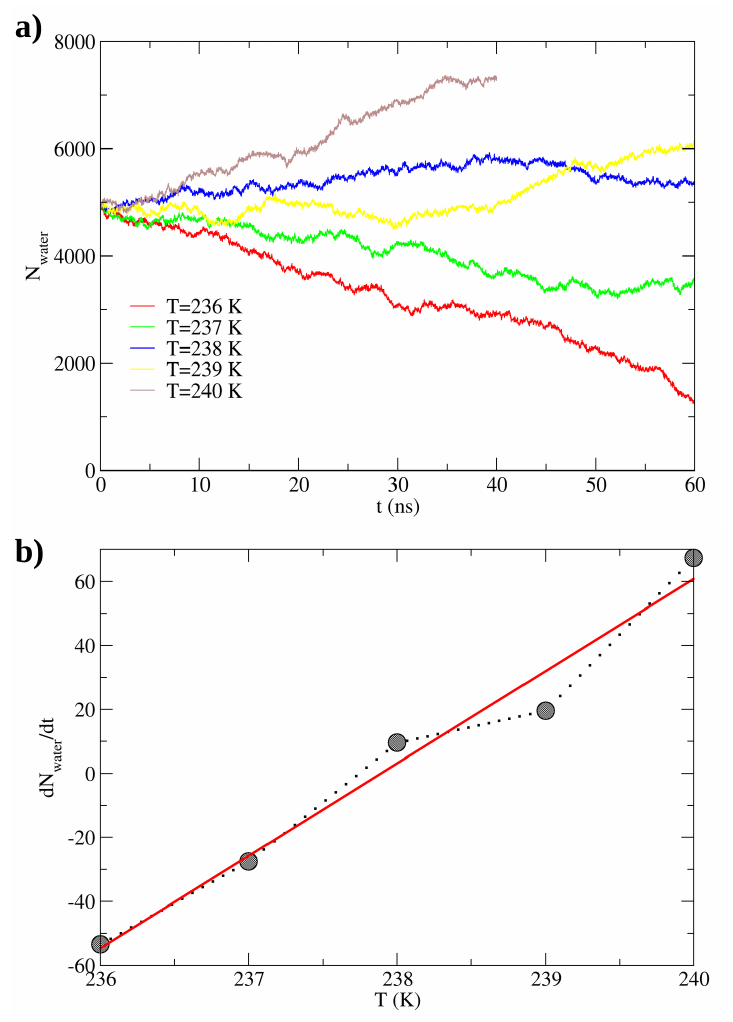}
    \caption{a) The evolution of number of liquid molecules in time 
    for different temperatures around the melting point. b) Slopes of the melting/freezing modes plotted as a function of temperature. }
    \label{fig:melt}
\end{figure}

A well-known limitation of point charge models is the difficulty to predict simultaneously both the melting point and the temperature of maximum density. TIP4P/2005, which is otherwise quite good as a model of liquid water, predicts the melting point at about 250~K. TIP4P/$\varepsilon$ is no exception, and previous reports of its melting point yield $T_m=240$~K. However, being aware of the uncertainties in the evaluation of the $T_m$ due to the finite-size effects \cite{conde17}, we have repeated the estimation of $T_m$ using the direct coexistence method with a large system size comprising of $10240$ water molecules in total. The secondary prism face (pII) of ice was exposed to water as it exhibits the fastest kinetics \cite{nada05} and  the evolution of a number of liquid molecules as a function of time was monitored for several different temperatures. This can be seen in Figure~\ref{fig:melt}-a.

We can estimate the  freezing/melting rates from the slope of the number of liquid molecules as a function of time. These are then plotted as a function of temperature, and the melting point is estimated as the interpolation of this curve to zero melting rate \cite{baran24b}.  The freezing/melting rates are plotted in Figure~\ref{fig:melt}-b. The intersection of these rates with the x-axis provides an estimated melting temperature of  $T_{m,\varepsilon}=237.9$~K. This value is smaller than the recently estimated melting point of the TIP4P/2005 model of $T_{m,2005}\approx250$~K \cite{conde17}. For point charge models with similar dipole moments, the location of the melting point correlates with the quadrupole moment, $Q_T$ \cite{abascal07}. As shown in Table~\ref{tab:models}, TIP4P/$\varepsilon$ has lower value of $Q_T$ than the TIP4P/2005, explaining the shift in melting point by about $12$~K. 

Once the melting temperature at atmospheric conditions has been evaluated, auxiliary bulk simulations at $T_m$ have been launched in the $NpT$ ensemble to estimate the coexistence densities of liquid and ice phases. Moreover, the melting enthalpy $\Delta H_{melt}$ has been extracted from these simulations. Using that value, the slope of the Clausius-Clapeyron equation, $dp/dT$, has also been estimated. These properties are all summarized in Table~\ref{tab:comparison}. 

\subsection{Transport properties}

Next, we consider  results for  self-diffusion coefficients. In Figure~\ref{fig:diffusion}-a a comparison between the TIP4P/2005 and TIP4P/$\varepsilon$ water model is shown for different temperatures at the atmospheric pressure of $p=1$~bar. For the latter force  field, the data are shown both with and without the YH correction. In Figure~\ref{fig:diffusion}-b a magnified region is presented. It is evident that the two models are equivalent once the YH correction is included in both cases. 

In Figure~\ref{fig:diffusion}-c, the relation of self-diffusion coefficients with pressure at three different temperatures is shown. Solid lines correspond to the experimental results. It is clear that the TIP4P/$\varepsilon$ is able to reproduce experimental data of the pressurized water in a wide range of thermodynamic conditions. 

\begin{figure}[h!]
    \centering
    \includegraphics[width=\linewidth]{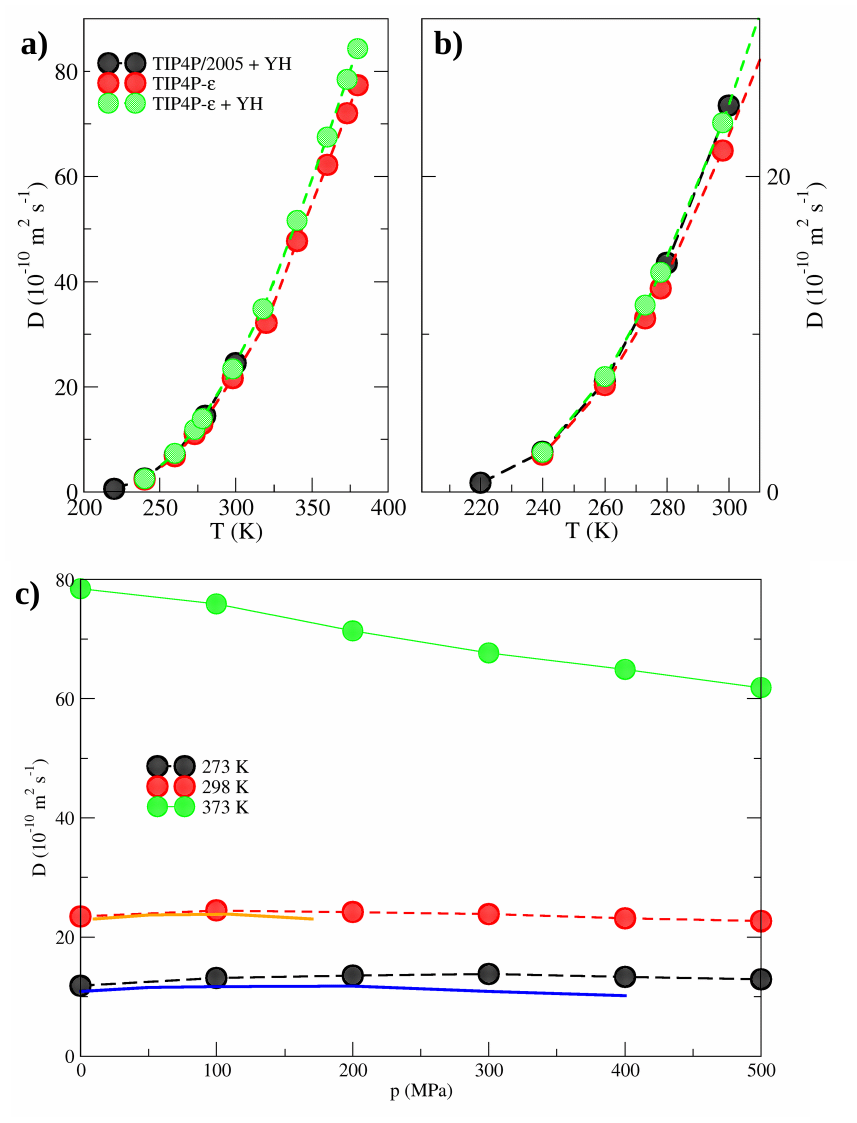}
    \caption{a) Relation of self-diffusion coefficient with the temperature at the atmospheric pressure $p=1$~bar for the TIP4P/$\varepsilon$ and TIP4P/2005 water models. The data for the latter is taken from Ref.~\cite{montero18}. b) magnified region of part a). c) Relation of self-diffusion coefficient with pressure at three different temperatures for the TIP4P/$\varepsilon$ water model. Experimental results is shown as a solid line. Data for $T=273$~K is taken from Ref.~\cite{prielmeier88} while for $T=298$~K from Ref.~\cite{krynicki78}.}
    \label{fig:diffusion}
\end{figure}

By  virtue of the Stokes-Einstein relation, it is expected that the shear viscosity values should also be equivalently well reproduced by the TIP4P/$\varepsilon$ water model as it is inherently coupled to the self-diffusion coefficient. Indeed, it is the case as can be seen in Figure~\ref{fig:visc-pure} where the results overlap with the data from Montero de Hijes \textit{et al.} \cite{montero18} for the TIP4P/2005 model. In Figure~\ref{fig:visc-pure}-b a magnified region is presented. 

In Figure~\ref{fig:visc-pure}-c, the relation of shear viscosity with pressure at three different temperatures is shown. Remarkably, the shear viscosities at $T=273$~K are around 10\% too small as compared to the experimental data, however, the trend in which the viscosity reaches its minimum and increases again with pressure is preserved. Nevertheless, this is also the case for the TIP4P/2005 water model as demonstrated in Figure~2 in Ref.~\cite{montero18}. As the temperature increases, the agreement with experimental results improves. 

\begin{figure}[h!]
    \centering
    \includegraphics[width=\linewidth]{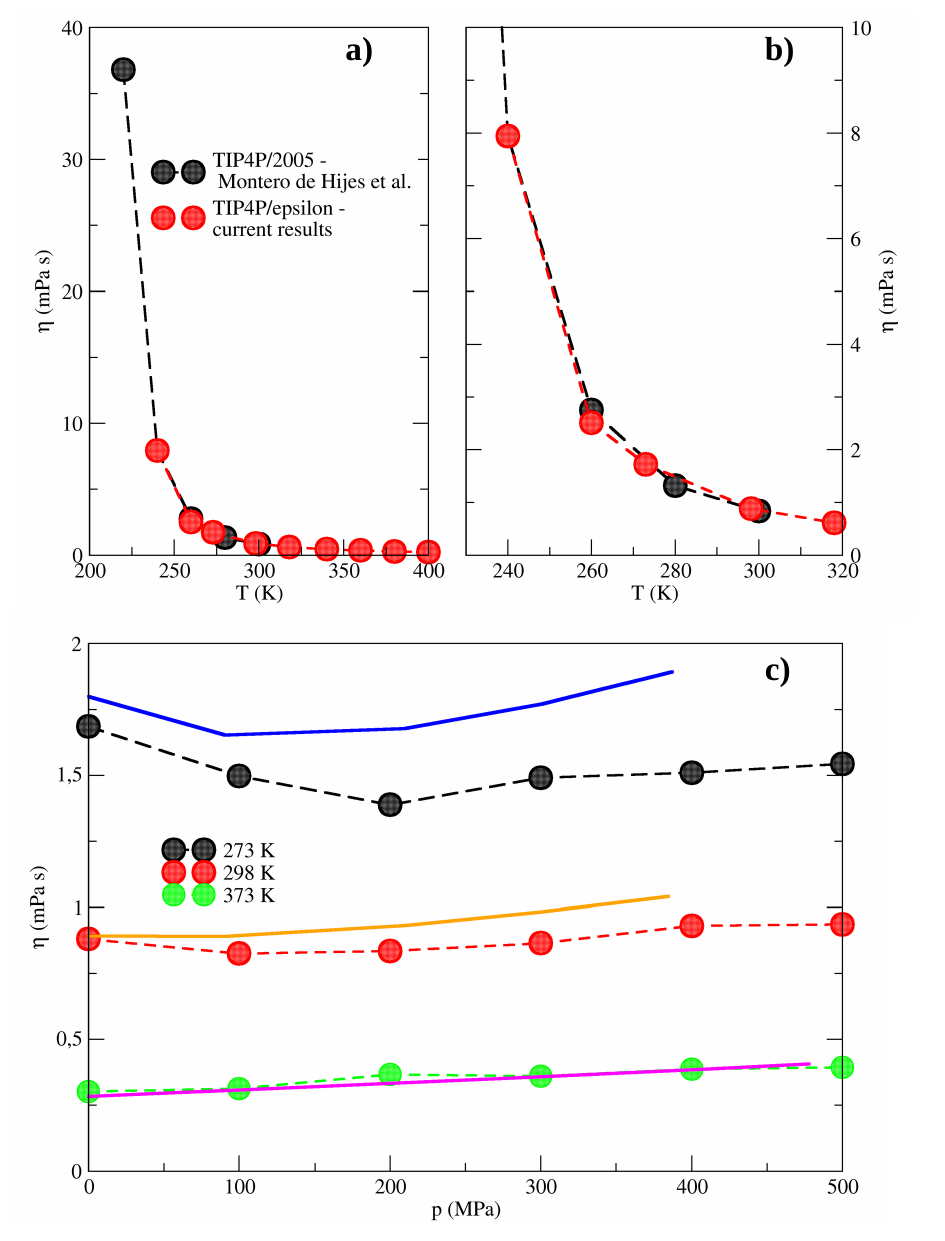}
    \caption{a) Relation of the shear viscosity with the temperature at the atmospheric pressure $p=1$~bar for the TIP4P/$\varepsilon$ and TIP4P/2005 water models. The data for the latter are taken from Ref.~\cite{montero18}. b) magnified region of part a). c) Relation of the shear viscosity model with pressure  for the TIP4P/$\varepsilon$ at three different temperatures. Experimental results are taken from Ref.~\cite{Harris04, Harlow67}}.
    \label{fig:visc-pure}
\end{figure}

\subsection{Comparison of the models}

So far, for selected properties, we have seen that the TIP4P/$\varepsilon$ water model is essentially identical to the TIP4P/2005 force field in a wide range of properties, with the exception that the former also reproduces experimental values of the dielectric constant. Therefore,  we are now in a position to perform a similar comparison as has been recently done by Sedano \textit{et al.} (TIP4P/2005 vs OPC models), using the VA-test method \cite{vega11}. 

The results for different properties are shown in Table~\ref{tab:comparison}. The comparison is done exactly as in Ref.~\cite{sedano24}, except for the heat capacities at constant pressure, which were not calculated in this work. All the results for the TIP4P/2005 and OPC water models are taken from that reference {\color{black} (except for the viscosity at 298 K which is taken from \cite{gonzalez10})}, whereas properties of the TIP4P/$\varepsilon$ are calculated by ourselves unless marked as an asterisk. In the latter case, the value for a given property is taken from Ref.~\cite{fuentes14}. The tolerances allowed for variations of properties in the rating scheme are as those reported in Table~2 of Ref.~\cite{vega11}.

 Despite the wide range of properties tested in the VA benchmark, both models perform  quite well. TIP4P/2005 is superior for the prediction of the melting and critical points, but is very similar to TIP4P/$\varepsilon$ for all other properties, including ice polymorph densities, liquid densities, shear viscosities and diffusion coefficients.  The overall performance is almost the same for both models (7.59 for TIP4P/2005 vs 7.54 for TIP4P/$\varepsilon$) while the OPC is clearly inferior (with a score of 6.26). As a bonus, 
TIP4P/$\varepsilon$ predicts accurately the dielectric constants, compared to TIP4P/2005 which provides results that are ca. 25~\% too small. 
Surprisingly, this large increase of $\epsilon$ is achieved with only a minor increase of the molecular dipole moment. The implication is that orientation correlations of the TIP4P/$\varepsilon$ model must be very different from those predicted by TIP4P/2005, yet the prediction of densities and transport properties in both models remain very similar. It would seem that changes in the Lennard-Jones parameters from one model to the other compensate somehow for the differences in charge distributions. The question then remains, whether one of the models is predicting the actual electric field about a water molecule better than the other, but this cannot be resolved from our results based on the properties of pure water.

\begin{table*}[h!]
    \centering
\resizebox{0.8\textwidth}{!}{
    \begin{tabular}{ccccc}
        Property & Exp. & TIP4P/2005  & TIP4P/$\varepsilon$  & OPC \\
        \hline
        \hline
        \multicolumn{5}{c}{Enthalpy of phase change (kJ/mol)} \\
        \hline
        $\Delta H_{melt}$ & 6.02 & 4.73 & 4.05 & 4.48 \\
        $\Delta H_{vap}$ & 44.02 & 50.17 & $53.24^*$ & 53.97 \\
        \hline
        \multicolumn{5}{c}{Critical properties} \\
        \hline
        $T_c$ (K) & 647.1 & 641.4 & $665^*$ & 697 \\
        $p_c$ (bar) & 220.64 & 146 & 135$^*$ & 168 \\
        $\rho_c$ (g/cm$^3$) & 0.322 & 0.31 & $0.32^*$ & 0.291 \\
        \hline
        \multicolumn{5}{c}{Surface Tension (mN/m)} \\
        \hline
        $\sigma_{300K}$ & 71.7 & 69.3 & 69$^*$ & 75.3 \\
        $\sigma_{450K}$ & 42.88 & 41.8 & 43.8$^*$ & 54.3 \\
        \hline
        \multicolumn{5}{c}{Melting properties} \\
        \hline
        $T_m$ (K) & 273.15 & 250 & 237.89 & 244.5 \\
        $\rho_l$ (g/cm$^3$) & 0.997 & 0.994 &  0.987 & 0.996 \\
        $\rho_s$ (g/cm$^3$)& 0.917 & 0.921 & 0.920 & 0.895 \\
        $dp/dT$ (bar/K)  & -140  & -132  & -130.26 & -90 \\
        \hline
        \multicolumn{5}{c}{Orthobaric densities (g/cm$^3$) and TMD (K)} \\
        \hline
        TMD & 277 & 277.3 & 276$^*$ & 270.1 \\
        $\rho_{298K}$ & 0.999 & 0.997 & 0.997 & 0.998 \\
        $\rho_{400K}$ & 0.9375 & 0.935 & 0.936 & 0.942 \\
        $\rho_{450K}$ & 0.8903 & 0.885 & 0.890 & 0.9 \\
        \hline
        \multicolumn{5}{c}{Isothermal compressibility ($10^{-6}$~bar$^{-1}$)} \\
        \hline
        $\kappa_T$~(1 bar, 298.15 K) & 45.3 & 46.4 & 45.6$^*$ & 44.5 \\
        $\kappa_T$~(1 bar, 360 K) &  47 & 50.9 & 49.1$^*$ & 44.4 \\
        \hline
        \multicolumn{5}{c}{$T_m-\mathrm{TMD}-T_c$ ratios} \\
        \hline
        $T_m$(Ih)/$T_c$ & 0.4221 & 0.3898 & 0.3571 & 0.3508 \\
        TMD/$T_c$ & 0.4281 & 0.4323 & $0.4150^*$ & 0.3875 \\
        TMD-$T_m$ & 3.85 & 27.3 & 38.52 & 25.6 \\
        \hline
        \multicolumn{5}{c}{Static dielectric constant} \\
        \hline
        $\varepsilon_r$~(liq, 298 K) & 78.5 & 57 & 79.16 & 78 \\
        \hline
        \multicolumn{5}{c}{Densities of ice polymorphs (g/cm$^3$)} \\
        \hline
        $\rho$~(Ih, 1 bar, 250 K) & 0.92 & 0.921 & 0.919 & 0.894 \\
        $\rho$~(II, 1 bar, 123 K) & 1.19 & 1.211 & 1.200 & 1.176 \\
        $\rho$~(V, 5.3 kbar, 223 K) & 1.283 & 1.272 & 1.268 & 1.239 \\
        $\rho$~(VI, 11 kbar, 225 K) & 1.373 & 1.369 & 1.373 & 1.335 \\
        \hline
        \multicolumn{5}{c}{EOS high pressure} \\
        \hline
         $\rho$~(10 kbar, 373 K) & 1.201 & 1.204 & 1.201 & 1.189\\
         $\rho$~(20 kbar, 373 K) & 1.322 & 1.321 & 1.317 & 1.299 \\
        \hline
        \multicolumn{5}{c}{Self-diffusion coefficient (cm$^2$/s)} \\
        \hline
        $\ln(D_{278K}) [D\times10^5]$ & -11.24 [1.31] & -11.23 [1.33] & -11.18 [1.40] & -11.14 [1.46] \\
        $\ln(D_{298K}) [D\times10^5]$ & -10.68 [2.30]& -10.67 [2.32] & -10.66 [2.34] & -10.65 [2.38]\\
        $\ln(D_{318K}) [D\times10^5]$ & -10.24 [3.57] & -10.25 [3.52] & -10.26 [3.49] & -10.25 [3.52] \\
        $E_a$ (kJ/mol) & 18.4 & 18 & 16.93 & 16.3 \\
        \hline
        \multicolumn{5}{c}{Shear viscosity (mPa$\cdot$s)} \\
        \hline
        $\eta$ (298 K) & 0.9 & 0.85 & 0.881 & 0.79 \\
        $\eta$ (373 K) & 0.28 & 0.28 & 0.293 & 0.3 
    \end{tabular}
    }
    \caption{Comparison between TIP4P/2005, TIP4P/$\varepsilon$, and OPC water models. The values of the reported quantities for TIP4P/2005 and OPC 
    are taken from \cite{sedano24} whereas for the TIP4P/$\varepsilon$ are calculated by us, unless marked by an asterisk which is then taken from \cite{fuentes14}.}
    \label{tab:comparison}
\end{table*}

\clearpage

\section{Electrolyte properties of TIP4P/$\varepsilon$ model}

In order to elucidate which of the two models is more consistent, it is required to test their performance beyond pure water properties. Particularly, a strategy to test which of the two charge distributions is electrostatically more significant is to test the models under strong electric fields, such as that generated locally in electrolyte solutions. {\em A priori}, one can tell that TIP4P/$\varepsilon$ could perform better for the solubilities of salts, as it predicts accurately the dielectric constant. However, if this agreement is at the cost of spoiling the force field, the remaining solution properties will not be predicted correctly.

\subsection{Joung-Cheatham NaCl model}

In the present work, we have decided to use the NaCl parameters of the force field that has been developed by Joung and Cheatham (JC) \cite{joung08}. The parameters used in conjunction with the TIP4P/$\varepsilon$ water model are shown in Table~\ref{tab:jc}. 

The motivation behind the choice of the JC model is three-fold. Firstly, the model has been parametrized to accurately reproduce the properties of the NaCl crystal and its melt. This constrains the parameters and avoids possible inconsistencies that could result from fitting to the solution properties alone. Secondly, as a corollary of the former comment, the JC model sets unit charges to the Na$^+$ and Cl$^-$ ions, which is the textbook expectation for charge distributions of dissolved ions. Thirdly, the model parameters have not been optimized for either TIP4P/2005 or TIP4P/$\epsilon$. This is convenient, as we can therefore check the performance of the water models without any {\em a priori} bias. The comparison will therefore allow us to check whether (i) the model electrolyte is transferable to other water models and (ii) reproducing the experimental values of  dielectric constant improves the results. 

\begin{table}[h!]
    \centering
    \begin{tabular}{ccccc}
        LJ Interaction & $\sigma$~(\AA) & $\varepsilon$~(kcal/mol) & Charge &  q (e)  \\
        \hline
        \hline
         Na$^+$-Na$^+$  &  2.160 & 1.4752 & Na$^+$ & $+1.0$ \\
         Cl$^-$-Cl$^-$  &  4.830 & 0.0536 & Cl$^-$ & $-1.0$\\
         Na$^+$-Cl$^-$  &  3.495 & 0.2811 &  \\
         Na$^+$-O       &  2.6625 & 1.0681 & \\
         Cl$^-$-O       &  3.9975 & 0.2035 & \\
         O-O            &  3.165  & 0.7732 &  O   & 0.0 \\
                        &         &         &  H   & $+0.527$ \\ 
                        &         &         &  M   & $-1.054$ \\
        \hline
        \hline
    \end{tabular}
    \caption{Joung-Cheatham NaCl parameters combined with TIP4P/$\varepsilon$ water model.
    The only difference for the TIP4P/2005 water model would be the (i) water charges and LJ interactions and (ii) NaCl-water cross interactions as shown in Table~II in Ref.\cite{benavides16}.}
    \label{tab:jc}
\end{table}

\subsection{Density as a function of molality}

The density of NaCl solutions as a function of molality is shown in
Figure~\ref{fig:molality}-a. It is evident that the JC model performs better with  TIP4P/$\varepsilon$ than with TIP4P/2005.  The discrepancies are the larger, the larger the concentration of the solution, but the differences are still within an acceptable margin. Both water models yield acceptable predictions. In this case, the full charges that might be too large for solvation in  TIP4P/2005 water can be better accommodated by the TIP4P/$\varepsilon$ model due to its larger dielectric constant. The better performance of this model in this particular case could be, nevertheless, accidental. But it is worth noting that the results are rather similar to predictions from the Madrid-2019 ion force field, which has scaled charges and Lennard-Jones parameters specifically parametrized for use with TIP4P/2005.

\begin{figure}[h!]
    \centering
    \includegraphics[width=\linewidth]{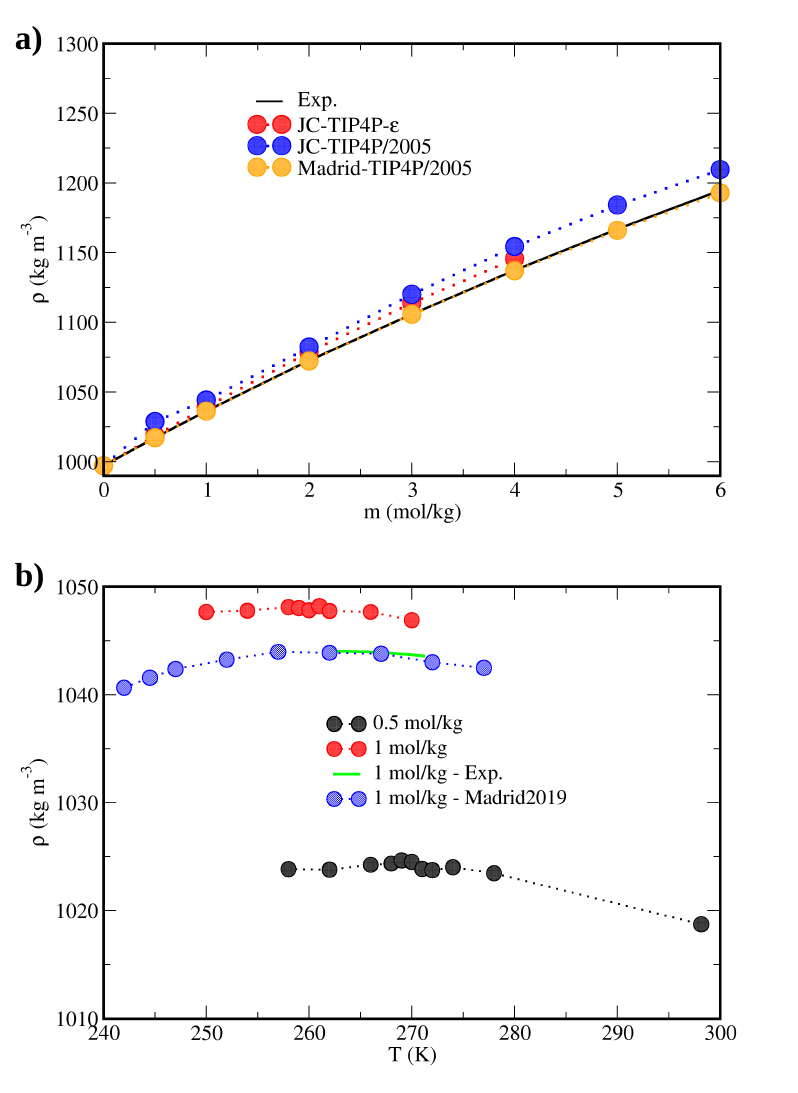}
    \caption{a) Density as a function of molality for the JC and Madrid-2019 NaCl
    force fields used in conjunction with TIP4P/2005 and TIP4P$/\varepsilon$ water models. 
    Solid line is the experimental data, taken from Ref.~\cite{pitzer84}.
    b) Density as a function of temperature for two different concentrations
    for the JC-TIP4P/$\varepsilon$ electrolyte solutions.{\color{black} Experimental data and for Madrid-2019 force field is taken from \cite{sedano22}} }
    \label{fig:molality}
\end{figure}

\subsection{Temperature of maximum density}

The TMD is quite sensitive to the addition of salts. Results for the density of JC salts of different concentrations dissolved in TIP4P/$\varepsilon$ water are shown in Figure~\ref{fig:molality}-b. From these plots, the temperature of maximum density (TMD) has been estimated as shown in Table~\ref{tab:tmd}, which also compares results for JC ions in TIP4P/2005 water obtained previously \cite{benavides16, benavides17}. It is clear that both models reproduce experimental results reasonably well at all considered NaCl concentrations, despite the use of elementary charges for the ions. {\color{black} The same conclusion can be drawn regarding the Madrid-2019 force field. However, the main advantage is that it not only captures the TMD but also reproduces the experimental densities within a wide temperature range which is expected based on the Figure~\ref{fig:molality}-a.}

\begin{table}[h!]
    \centering
    \begin{tabular}{ccccc}
        m (mol/kg) & Exp. & JC-TIP4P/2005 & JC-TIP4P/$\varepsilon$ & Madrid-2019\\
        \hline
        \hline
         0 &   277.1 & 278 & 276  & 278 \\
         0.5 & 270.5 & 267 & 269  & \\
         1.0 & 263.5 & 259 & 258/261 & 260.7 \\
    \end{tabular}
    \caption{Temperatures of maximum density (in K) at $p=0.1$~MPa for different
solutions of NaCl in water as obtained from simulations and experiments. The results for the TIP4P2005 are taken from Ref.~\cite{benavides16, benavides17}, {\color{black}for the Madrid-2019 from \cite{sedano22}}, and experimental results from Ref.~\cite{washburn28}.
}
    \label{tab:tmd}
\end{table}

\subsection{{\color{black}Solvent} dielectric constant as a function of molality}

Dielectric properties are also quite sensitive to salt concentration, so it is interesting to test how this dependence is described by TIP4P/$\varepsilon$, which already exhibits a good dielectric constant for pure water. {\color{black} However, one needs to bear in mind that due to the conducting nature of electrolyte solutions, the dielectric constant measured by Eq.~\ref{eq:epsilon}, which holds for an insulating media, is not directly accessible from the experiments \cite{hubbard79, cruz18, cruz20}. In the latter case, the dielectric constant is measured by extrapolation to zero frequency after the removal of the divergent conductive contributions. However, the resulting  property peaks up additional  dynamical contributions  related to cross correlations between the dipole moment and the charge current \cite{caillol86} that are not contained in Eq.~\ref{eq:epsilon}. Although this contribution appears to be relatively small,\cite{caillol86} a direct comparison with experimental data shown should be taken with some caution.}

Figure~\ref{fig:eps}-a. shows that TIP4P/$\varepsilon$ can predict very accurately a {\color{black} sharp drop} of the dielectric constant with salt concentration, providing quantitative results for up to two molal concentrations. Results for larger concentrations deteriorate but remain quite reasonable up to the measured concentrations of about six molal.

A similar performance for the TIP4P/2005 model can obviously not be expected, since the dielectric constant is predicted already 25\% too low for pure water. However, it is interesting to note that the strong decrease in the dielectric constant is properly predicted by both models. To see this, Figure~\ref{fig:eps}-b displays the relative drop of the dielectric constant, $\Delta \epsilon/\epsilon=(\epsilon(C) - \epsilon(C=0))/\epsilon(C=0)$. In this scale, both models behave similarly. TIP4P/$\varepsilon$ predicts to sharp a drop of the  relative dielectric constant, while TIP4P/2005 appears to predict values somewhat larger than found in the experiment.

\begin{figure}[h!]
    \centering
    \includegraphics[width=\linewidth]{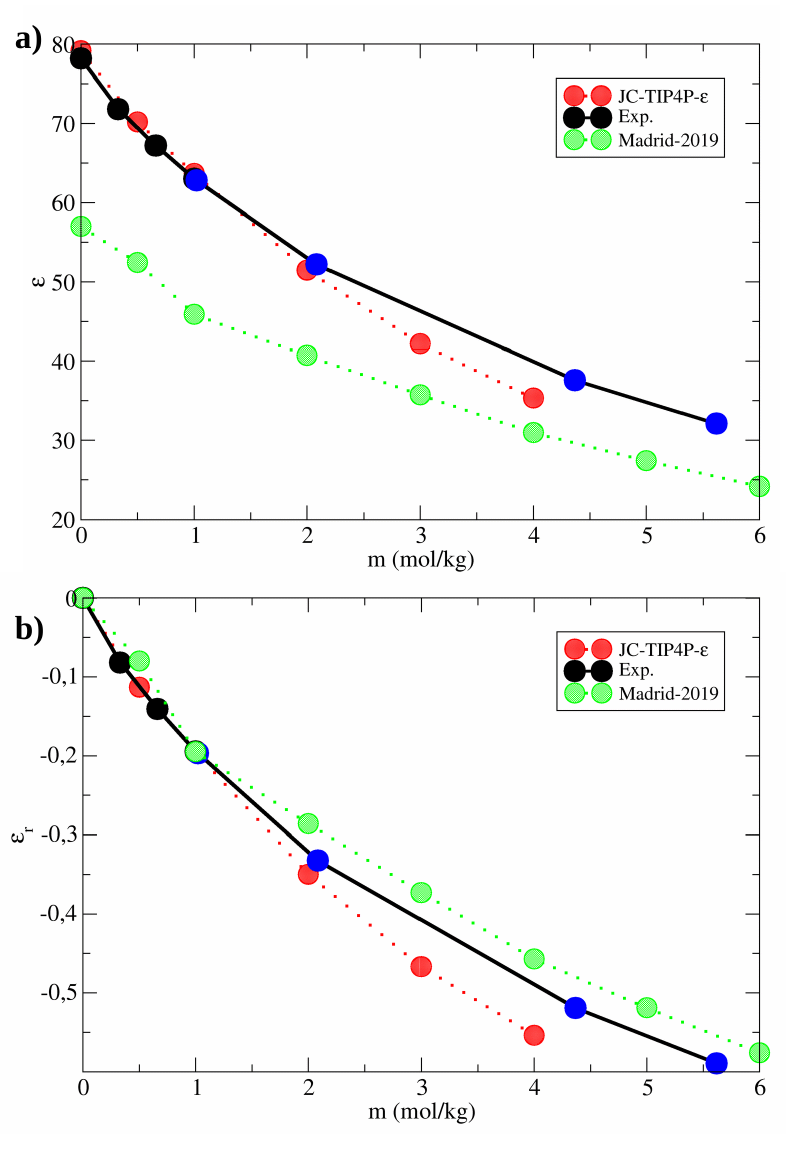}
    \caption{{\color{black} Solvent} dielectric constant (a) and relative dielectric constant (b) as a function of NaCl molality 
    at $T=298.15$~K. 
    The experimental data are taken from \cite{Haggis52} (black circles)
    and \cite{christensen66} (blue circles).  }
    \label{fig:eps}
\end{figure}

\subsection{Viscosity as a function of molality}

\begin{figure}[h!]
    \centering
    \includegraphics[width=\linewidth]{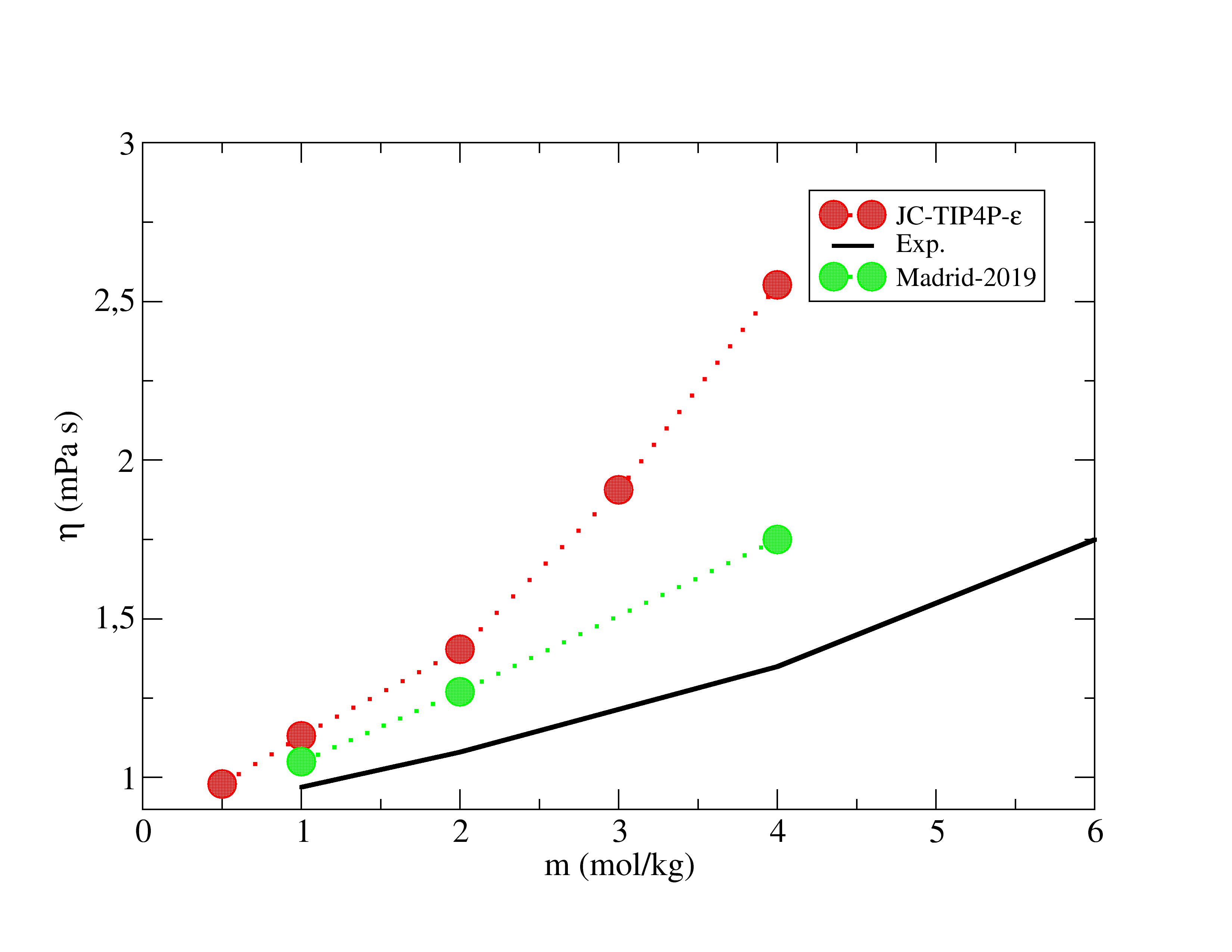}
    \caption{Viscosity as a function of molality for NaCl solution. 
    The solid line shows experimental results}
    \label{fig:visc}
\end{figure}

As a final check, we now test the role of salt concentration on the viscosity of the solutions in Figure~\ref{fig:visc}. Unfortunately, in this case, the results for the JC ions in combination with the TIP4P/$\varepsilon$ are very disappointing, with predicted viscosities that are far too large  
in the entire concentration range studied. This has been observed previously for the dynamic properties of electrolyte solutions of rigid point charge solvents. The large charge of the ions leads to trapping of the water molecules, which have great difficulty escaping from the solvation cage for lack of local charge fluctuations in the rigid point charge models \cite{blazquez23a}. 
This explains why the Madrid-2019 force field, with ionic charges scaled to 0.85 times the elementary charge performs significantly better. Blazquez \textit{et al.} \cite{blazquez23a} have shown that the discrepancy can be removed by further scaling the charges to even  smaller values (equal to $\pm 0.75$), but unfortunately, this is not satisfactory either, as other properties are then not accurately predicted.

\subsection{Melt and lattice properties}

Prediction of the thermodynamic properties of NaCl aqueous solutions has been demonstrated in previous subsections. 
It was shown that JC NaCl dissolved in TIP4P/$\varepsilon$ water predicts solution properties rather accurately without any additional parametrization. 

Since the JC model was parametrized with the NaCl crystal and melt as a target, the properties of the melt and the solid at the melting point are also rather accurately predicted, as shown in Table~\ref{tab:melt}. On the other hand, the Madrid-2019 force field is somewhat less accurate, as this set of properties was sacrificed for better performance of solution properties.  

Overall, this means that TIP4P/$\epsilon$ appears as a rather robust model for the study of solution properties, as it provides a rather fair account of solution properties for models parametrized for the pure salt properties. This makes the model transferable to a significant extent and is expected to provide an accurate account of the phase coexistence of aqueous electrolytes.

\begin{table}
    \centering
    \begin{tabular}{c|ccc}
    \hline
    \hline
        Property & Exp.  & JC-TIP4P/$\varepsilon$  & Madrid-2019 \\
        $\rho_{melt}$ (kg/m$^3$) at $T=1073.8$~K & 1556 & 1410  & 1331\\
         $\rho_{solid}$ (kg/m$^3$) at $T=298.15$~K & 2011 & 2165  & 2050 \\
         $E_{lattice}$ (kJ/mol) & 786 & 785.5 & 607 \\
    \hline
    \hline
    \end{tabular}
    \caption{Densities of the molten and anhydrous NaCl at $p=1$~bar and experimental melting temperature $T_m=1073.8$~K and $T=298.15$~K, respectively. Lattice energy has been estimated at the same conditions as the density.}
    \label{tab:melt}
\end{table}

\section{Conclusions}

In this work, we have considered the role of the dielectric constant as a parametrization target for accurate empirical force fields. Based on previous results, including some of our own work, we had expected that an accurate account of the dielectric constant of simple point charge models could be at the cost of compromising the accuracy of a large number of properties \cite{macdowell10,aragones11b,sedano24}. A dramatic example is the case of the SPC/E model, which predicts a much better dielectric constant than most TIP4P models, but is unable to provide a qualitatively correct description of the solid phases \cite{sanz04}. A more recent example is the work of Sedano et al., who showed that OPC, a well-known model parametrized with the dielectric constant, performs significantly less well than TIP4P/2005 as based on the Vega-Abascal benchmark \cite{sedano24}.

These observations can be rationalized to some extent by the scaled-charge paradigm, which dictates that the nominal point charges of empirical force fields are screened by the electronic polarization of the surrounding medium \cite{leontyev09,leontyev11}. An implication is that the charges that dictate the force field are scaled versions of the charges that dictate the dipole moment \cite{macdowell10,aragones11b}. Whence, one expects that targeting the point charges to the dielectric constant will result in too large charges and dipole moments for the force field.

Contrary to these expectations, our results show that the TIP4P/$\varepsilon$ model, which has been parametrized to predict accurate densities and dielectric constants, can also provide a rather accurate account of many other properties with similar overall accuracy as the acclaimed TIP4P/2005 model. Indeed, we find that TIP4P/$\varepsilon$ can predict bulk phase properties (transport properties, thermal expansion coefficients, densities) of both liquid water and ice polymorphs, as well as surface tensions, with accuracy very similar to TIP4P/2005, while performing very well for dielectric constants over a wide range of pressures and temperatures. On the other hand, TIP4P/2005 provides a better description of phase boundaries, including liquid-vapor and freezing transitions.  Overall, these two models appear to perform almost identically on the conventional benchmark \cite{vega11} while significantly outperforming the OPC force field.

However, we find that the use of the dielectric constant as a target property has the advantage of enhancing model transferability  for the study of aqueous solutions. Indeed, our results show that TIP4P/$\varepsilon$ provides accurate predictions for solution densities and dielectric constants when used to dissolve model ions targeted to their crystal and melt properties. Particularly, this is achieved without the need to rescale charges, modify the Lorentz-Berthelot rule or tune the ion's Lennard-Jones parameters.  TIP4P/$\varepsilon$ thus appears as a very robust model for preliminary studies of solution properties when no additional parametrization can be afforded.

Unfortunately, the good description of solution thermodynamics of TIP4P/$\varepsilon$  breaks down when one seeks an accurate account of transport properties, a problem that is also shared by the TIP4P/2005 model \cite{blazquez23a}. This points to the limitations of point charge models and the need to address the problem of polarization in a physically meaningful way for next generation force fields with enhanced transferability. Current advances in computer architecture might well remedy the  computational overhead and make such models competitive in the very near future. In the meantime, TIP4P/2005 and TIP4P/$\varepsilon$ force fields appear as the two equally valid point charge models for use in routine calculations.

An intriguing question that remains unsolved is which of these two models is a better caricature of the actual charge distribution and local electric field created by a water molecule in the condensed phase. However, because this is actually a collective property that depends on many body interactions, it might simply be an ill-posed question. Be as it may, calculations of the dielectric constant of these models indicate that they predict a rather different Kirkwood $g_k$ factor, and accordingly, significantly different orientation correlations in the liquid phase. Measuring such correlations experimentally appears difficult, but it would allow us to discriminate between these two seemingly different empirical force fields. {\color{black} In the time being,  our results illustrate that a model targeted to reproduce dielectric constants can perform as well as TIP4P/2005  for pure water, and allows for an improved transferability in solution. In this sense, it seems that a good parametrization can be afforded without invoking the otherwise appealing concept of charge scaling.}

\section*{Acknowledgments}
We would like to thank Eva G. Noya and Enrique Lomba for helpful discussions. LGM wishes to thank his thesis advisor Carlos Vega for his guidance and friendship during his PhD. \L B would like to thank the Polish National Agency for Academic Exchange, under Grant No. BPN/BEK/2023/1/00006 Bekker 2023.  CDAG acknowledges the predoctoral fellowship "Contratos Predoctorales de Personal Investigador en Formación" from the Universidad Complutense de Madrid (CT15/23) and additional funding from Ministerio de Ciencia, Innovación y Universidades (MICIU), under grant FPU22/00869. LGM  would like to thank financial support from the Agencia Estatal de Investigaci\'on (Ministerio de Ciencia y Econom{\'i}a) under grant PID2023-151751NB-I00,
 funded by MCIN/AEI/10.13039/501100011033.  
 We also benefited from generous allocation of computer time at the Academic Supercomputer Center (CI TASK) in Gdansk.

\section*{Data Availability Statement}

The data that support the findings of this study are available from the corresponding author upon reasonable request.

\clearpage

\bibliography{biblio}

\end{document}